# A NUMERICAL MODEL TO PREDICT UNSTEADY CAVITATING FLOW BEHAVIOUR IN INDUCER BLADE CASCADES

O. Coutier-Delgosha[*]
coutier@enstay.ensta.fr

J. Perrin
jerome.perrin@hmg.inpg.fr

R Fortes Patella
fortes@hmg.inpg.fr

J.L. Reboud[**]
jean-luc.reboud@grenoble.cnrs.fr

*LEGI – INPG; BP 53 - 38041 Grenoble Cedex 9 – France*

[*] now at ENSTA – UER de Mécanique, Chemin de la Hunière, 91761 Palaiseau Cedex, France

[**] now at CNRS-LEMD, University of Grenoble, France.

## ABSTRACT

The cavitation behaviour of a four-blade rocket engine turbopump inducer is simulated. A 2D numerical model of unsteady cavitation was applied to a blade cascade drawn from the inducer geometry. The physical model is based on a homogeneous approach of cavitation, coupled with a barotropic state law for the liquid/vapour mixture. The numerical resolution uses a pressure-correction method derived from the SIMPLE algorithm and a finite volume discretization. Unsteady behaviour of sheet cavities attached to the blade suction side depends on the flow rate and cavitation number. Two different unstable configurations of rotating cavitation, respectively sub-synchronous and super-synchronous, are identified. The mechanisms that are responsible for these unstable behaviours are discussed, and the stress fluctuations induced on the blade by the rotating cavitation are estimated.

## INTRODUCTION

To achieve operation at high rotational speed and low inlet pressure, rocket engine turbopumps are generally equipped with an axial inducer stage. Under such operating conditions, cavitation develops on suction side of the blades and at inducer periphery near the tip. When pressure is decreased from cavitation inception, vapour develops more and more and finally leads to the inducer performance breakdown (Fig. 1).

Between low cavitating conditions and the performance drop, for some particular range of the cavitation number, unsteady phenomena may appear, associated with different blade cavitation patterns. Fig. 2 illustrates successive flow patterns observed experimentally in the case of the H2 inducer of Vulcain for different values of the cavitation parameter $\tau'$ [de Bernardi et al., 1993].

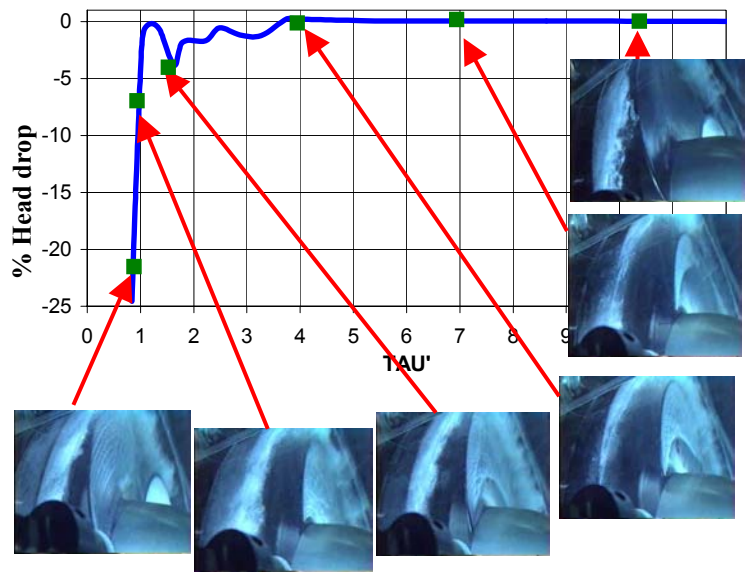

**Fig. 1.** Experimental performance chart for the H2 turbopump inducer of the Vulcain Engine (Ariane V). [Coutier-Delgosha et al., 2001] for the chart, and [Joussellin et al., 1998] for the pictures

At cavitation inception, a steady and balanced flow pattern with one short attached cavity on each blade is observed from flow visualizations (Fig. 2.1). When the cavitation parameter is slightly decreased, a steady and alternate cavitating configuration appears (only on four-blade inducers) with alternatively one short and one long cavity (Fig. 2.2). Then flow visualizations achieve to identify an unsteady flow pattern (Fig. 2.3): rotating cavitation appears at low cavitation parameter [Kamijo et al., 1977] just above breakdown. Unbalanced attached cavities are observed in the different channels, their distribution rotating faster than the inducer [de Bernardi et al., 1993, Pagnier et al., 1995] and leading to large radials loads on the shaft [Goirand et al., 1992]. Finally near the breakdown of the inducer, a steady and balanced flow pattern with fully developed cavitation is observed (Fig. 2.5).

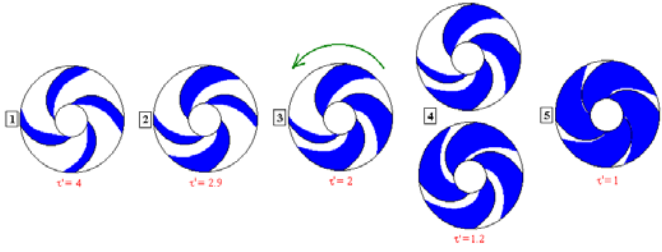

**Fig. 2.** Sketches of cavitation patterns for various cavitation parameters and their correspondence to the performance curve [de Bernardi et al., 1993].

Over the last few years, numerical models have been developed to predict the cavitation instability in inducers [Tsujimoto, 2001] These models are based on stability analyses and linear approach, taking into account the total flow rate variations through a cavitating blade-to-blade channel [Tsujimoto et al., 1993, Joussellin & de Bernardi, 1994, Horigushi et al., 2000a and b], or calculating the flow around attached cavities [Pilipenko et al., 1995, Watanabe et al., 1998 and 1999].

To improve the understanding and the prediction capability of cavitation instabilities, numerical and experimental analyses are developed through collaborations between the LEGI, the Rocket Engine Division of SNECMA Moteurs and the French space agency CNES. In the present paper, the cavitation behaviour of a four-blade inducer is simulated using a 2D model of unsteady cavitating flow developed at LEGI [Reboud & Delannoy, 1994, Coutier-Delgosha et al., 2000].

## NOMENCLATURE

| | |
|---|---|
| $Q, Q_n$ | Mass flow rate, nominal mass flow rate (kg/s) |
| $\mathbf{C}(Cm,Cu)$ | Velocity vector in fixed frame: $Cm=Q/(\rho S_{flow})$ |
| Lref | Reference length $Lref=\pi R_c/2$ |
| p, P | Static pressure, total pressure: $P = p + 0.5\, \rho_l\, \mathbf{C}^2$ |
| $p_v$ | Vapour pressure |
| $R_c$ | Cut radius for the passage from 3D to 2D |
| R, θ, Z | Cylindrical coordinates |
| Tref | Reference time: $Tref=Lref/Vref$ |
| Vref | Reference velocity: $Vref= R_c \Omega$ |
| $\Omega$ | Angular rotation speed of the inducer |
| U | Training velocity at inducer tip radius $R_{tip}$: $U=R_{tip}\,\Omega$ |
| τ, τ' | Cavitation parameter for inducers, $\tau'=\tau/\tau_c$, $\tau_c$ critical value. $\tau = (P_{inlet}-p_v) / (0.5\, \rho_l\, R_{tip}^2\Omega^2)$ |
| σ | Cavitation nb (downstream): $\sigma=(P_{outlet}-p_v)/(\tfrac{1}{2}\rho_l\, V_{ref}^2)$ |
| $f, f_0$ | Frequency, inducer rotation frequency |
| Ψ | Head coefficient $(P_{outlet}-P_{inlet})/(\rho_l\, R_{tip}^2\Omega^2)$ |
| $\rho_l, \rho_v, \rho$ | Density of the liquid, of the vapour, of the mixture |
| α | Local void ratio: $\alpha = (\rho-\rho_l) / (\rho_v-\rho_l)$ |

## 1. PHYSICAL AND NUMERICAL MODEL

Numerical simulation of the cavitating flow in the inducer was performed with the objective to take into account the cavitation sheets attached to the blades and their unsteady behaviour. The 2D numerical model of unsteady cavitating flow "IZ", developed in previous studies with the support of the CNES-Centre National d'Etudes Spatiales [Delannoy & Kueny, 1990, Reboud et al. 1998, Coutier-Delgosha et al., 2003], was adapted to 2D blade cascades [Coutier-Delgosha et al. 2000, Courtot, 2000]. The main features of the models are:

- The liquid-vapour mixture is described as a single fluid, whose density ρ rapidly varies between the pure liquid density $\rho_l$ and the pure vapour one $\rho_v$ when the static pressure in the flow field reaches the vapour pressure. The fluid density is managed by a barotropic state law ρ(p), as in [Coutier-Delgosha et al., 2001].
- 2D unsteady Reynolds averaged Navier-Stokes equations are applied to that single fluid with variable density. The numerical resolution uses a pressure-correction method derived from the SIMPLE algorithm and a finite volume discretization. The turbulence model used is the classical k-ε RNG model, associated with wall functions along solid boundaries.
- Other boundary conditions are (Fig. 5): imposed velocity at the mesh inlet (the flow velocity is deduced from the classical value $Q/\rho S_{flow}$), imposed static pressure at the outlet, and periodicity or connection conditions between the different channels of the blade cascade. These conditions are applied along non-matching boundaries, through a special treatment developed in [Coutier-Delgosha et al., 2000] to assure the best transfer of both mass and momentum fluxes.

## 2. GEOMETRY

The 3D inducer geometry considered in the present study is illustrated by the Fig. 3. Even at nominal flow rate, the inducer is designed to operate with non-zero incidence angle. Nevertheless, in the range of running conditions, that angle of attack remains very small, because of the variable thread and the associated curvature of the blades. In these conditions, the inlet back-flow is not relevant and a 2D geometry approach can be adopted.

Hence, to study the unsteady behaviour of the cavitating flow, some computations are performed in a four-blade cascade representing an entire inducer. The transformation from the 3D geometry to the 2D blade cascade leads to neglect the peripheral cavitation in the inducer: only attached sheets of cavitation on the blades will be considered. To obtain the 2D geometry, the 3D inducer (Fig. 3) is cut at a constant radius $R_c$.

Two values have been tested for $R_c$, namely $R_c$ = 70% of the tip radius R (cut m) or at the periphery of the runner (cut p). The resulting shape of the blade cut in the plane (Z/R, θ) is given in Fig. 4 in the two cases. In the present paper, only results concerning "cut m" are illustrated.

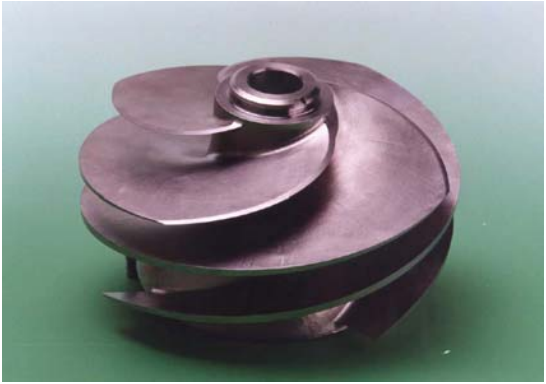

**Fig. 3.** 3D inducer

We use a 195x40 structured mesh per channel, giving a total of 31200 internal nodes when calculating the four-blade cascade. A special contraction of the mesh is applied close to the blades (Fig. 5) to ensure the size of the first cells between y+=30 and y+=100, and apply the k-ε RNG turbulence model with wall functions.

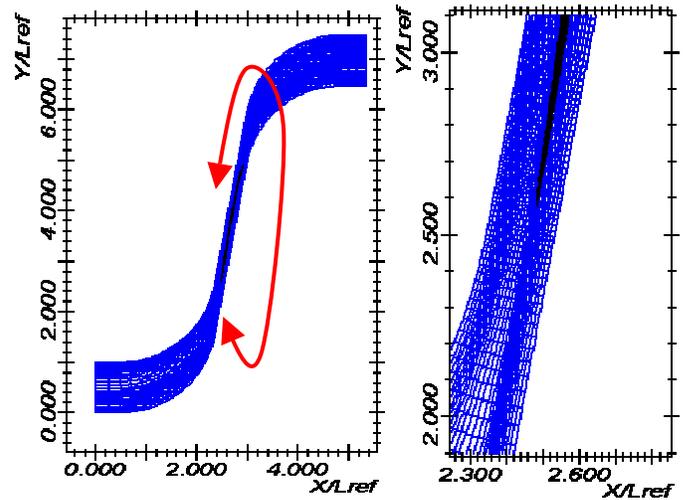

**Fig. 5.** Curvilinear, orthogonal mesh of a single channel (general view and zoom at the blade leading edge

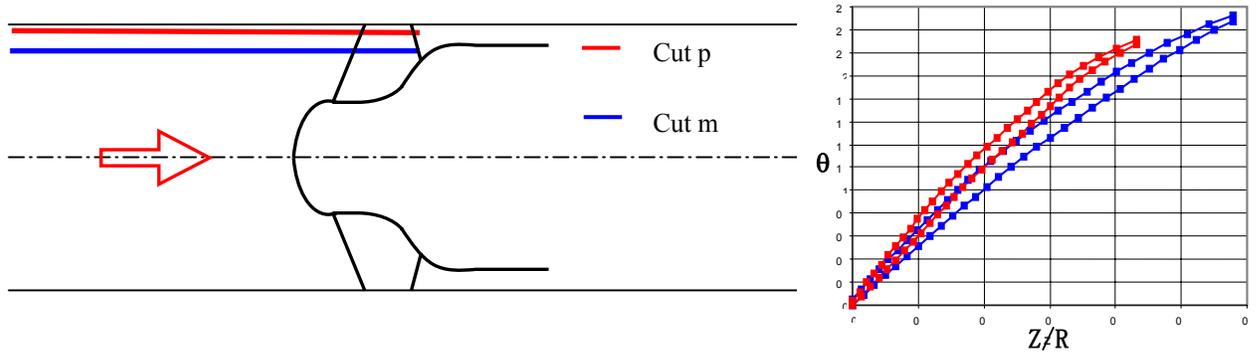

**Fig. 4.** Cut m and p to obtain the blade cascade

## 3. RESULTS

Quasi-steady calculations are performed on the one-channel mesh and unsteady computations are performed on the four-blade mesh. The periodicity condition is then applied between the 4th and 1st channels, as is the real runner. The time step, mesh and turbulence model are chosen to put the attention on the low frequency fluctuations of the attached cavity, more than to the local unsteadiness in the cavitation sheet wake (cloud shedding): the time step if fixed equal to 10% of the blade passage time $T_{ref}$. Successive time accurate computations are performed at fixed cavitation number and flow rate coefficient.

### 3.1 Head drop chart

Several four-channel computations are performed at nominal flow rate, with a cavitation number σ varying from low cavitating conditions down to the final performance drop. The corresponding head drop chart is drawn on Fig. 6 in blue. Some difference with the performance chart obtained with a single channel computation, represented in red, appears clearly, particularly for 0.6 < σ < 0.75. This is due to the appearance at these operating points of unstable cavitating behaviours. We first present in section 3.2 the results obtained in stable configurations (σ > 0.75 or σ < 0.6) and then the mechanisms of unstable configurations will be detailed in section 3.3.

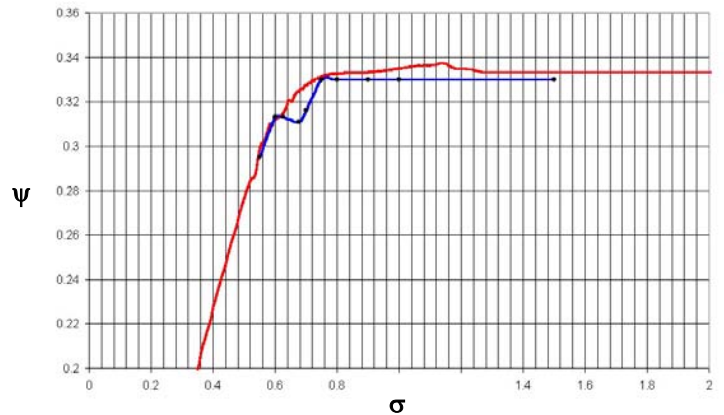

**Fig. 6.** Comparison between the performance charts obtained with four-blade computations (in blue) and with single blade computations (in red) at nominal flow rate.

## 3.2 Stable configurations

At nominal flow rate, stable configurations correspond to cavitation numbers higher than 0.75 or lower than 0.6. Between these two extrema, unstable conditions are obtained, which will be detailed in section 3.3.

For $\sigma = 0.75$, sheets of cavitation are still small, as indicated on Fig. 7a. The four cavities remain identical, even if the calculation is continued for a long time, or if perturbations are simulated to enhance a possible instability. The time evolution of the cavity length in the first channel, reported in Fig. 7b., confirms that sheets of cavitation are completely stable after the initial transient. For these flow conditions, the performance of the runner is only slightly affected by the presence of vapour, since its decrease does not exceed 5%.

A similar configuration is obtained for a cavitation number equal to 0.6, i.e. the inferior limit value. The sheets of cavitation are much more developed, and they are responsible for the important performance drop that is observed at this operating point: the obstruction due to the cavities modifies the velocity fields on the blades, which leads to the head drop.

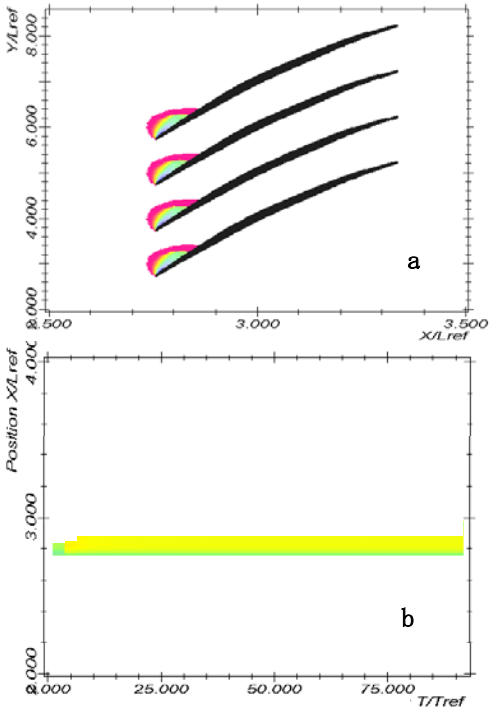

**Fig. 7.** Cavitation development for $\sigma = 0.75$ and $Q = Q_n$

a. Density fields (red corresponds to almost pure liquid, and dark blue to pure vapour (scale ratio x/y ≈10)
b. Time evolution of the cavity length in the first channel

It can be observed in Fig. 7. that the sheets of cavitation are mainly composed of liquid. As a matter of fact, the void ratio in the cavities remains always rather low: it does not exceed 60%, even for $\sigma = 0.6$.

## 3.3 Unstable configurations

For some particular values of the cavitation number, non-symmetrical configurations of sheets of cavitation on the four blades are obtained. These configurations are periodical, and they involve physical fluctuations of almost constant magnitude in the flow field, after the initial transient period. Two main configurations were observed:

### 3.3.1 First configuration for $\sigma = 0.7$ (nominal flow rate)

At this operating point, mass flow rates in the four blade-to-blade channels become different. Their evolution is reported in Fig. 8. It appears that at a given time, the mass flow rate in channels 1 and 3 is very different from the mass flow rate in channels 2 and 4. All evolutions are periodical, and suggest a coupling between respectively channels 1 and 3, and channels 2 and 4. The corresponding evolution of the four sheets of cavitation between T/Tref = 110 and T/Tref = 135 is represented in Fig. 9.

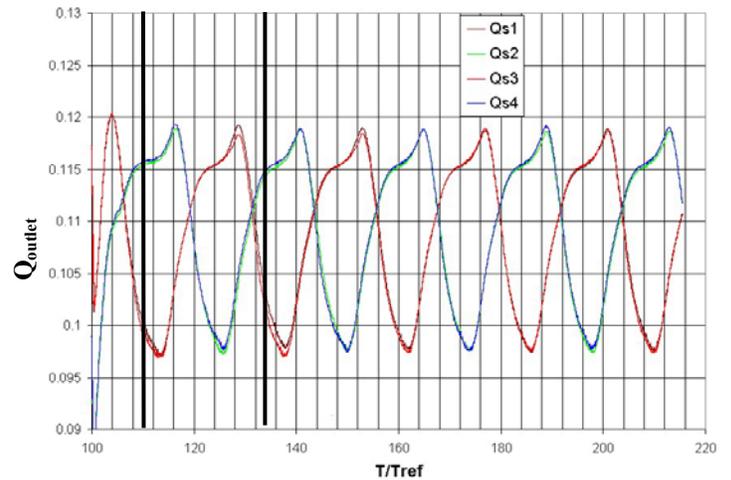

**Fig. 8.** Time evolution of the mass flow rate in the four channels ($\sigma = 0.7$ and $Q = Q_n$)

It is clear on Fig. 9 that mass flow rate variations in the different channels are strongly correlated with the size evolution of the sheets of cavitation. This continuous variation of cavities in channels 1 and 3 is the opposite of the one in channels 2 and 4, as for the mass flow rate in Fig. 8. In the first two configurations of Fig. 9, for example, sheets of cavitation are growing in channels 1 and 3, while they decrease in channels 2 and 4; at the same time, mass flow rate is decreasing in channels 1 and 3, and increasing in channels 2 and 4. This is due to the obstruction generated by the biggest cavities, which leads to a deviation of the flow towards free channels (i.e. channels in which cavities are small). This mechanism is amplified by the increase of the void ratio in the cavitation sheets during the step of growth: the vapour makes the flow curve round the leading edge of the corresponding blades, and a part of the flow rate passes in the upper channel. This is illustrated by Fig. 10, which shows (inside the blue circles) the deviation imposed by the biggest cavities to the flow.

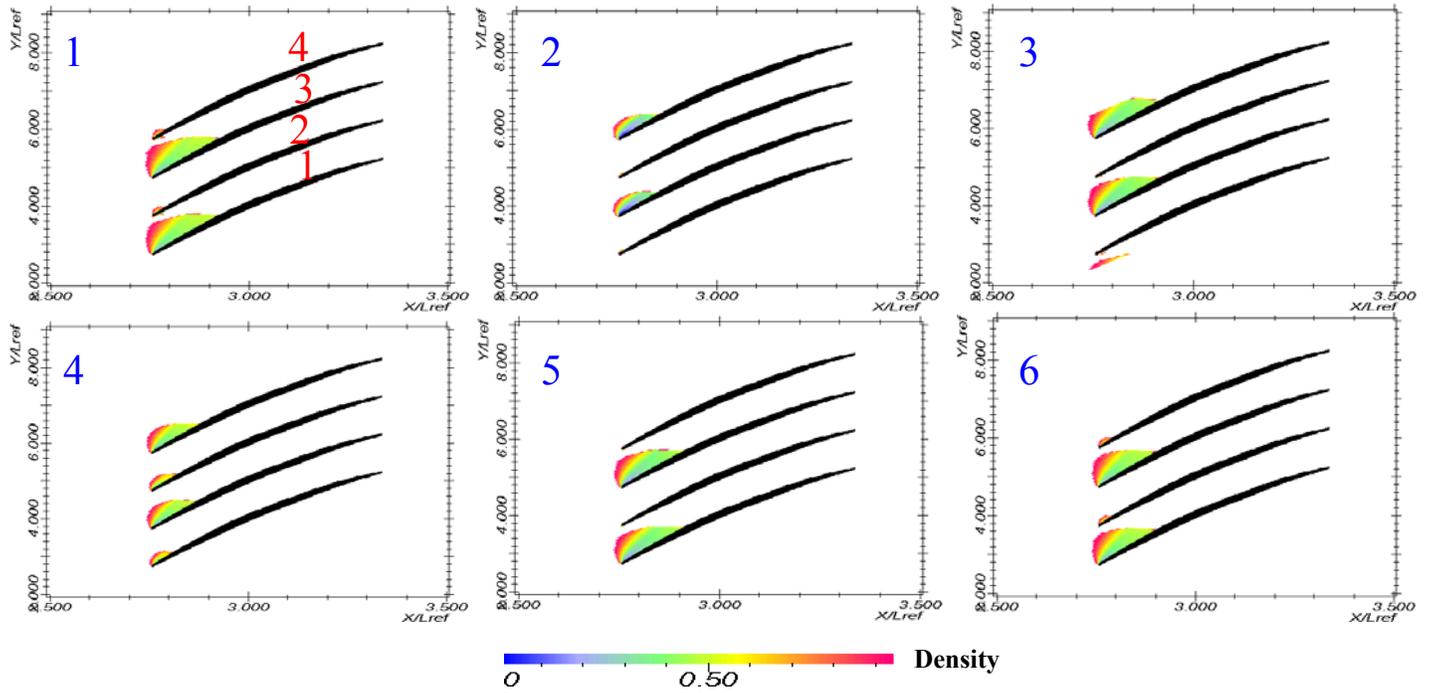

**Fig. 9.** Time evolution of the sheets of cavitation ($110 < T/T_{ref} < 135$)

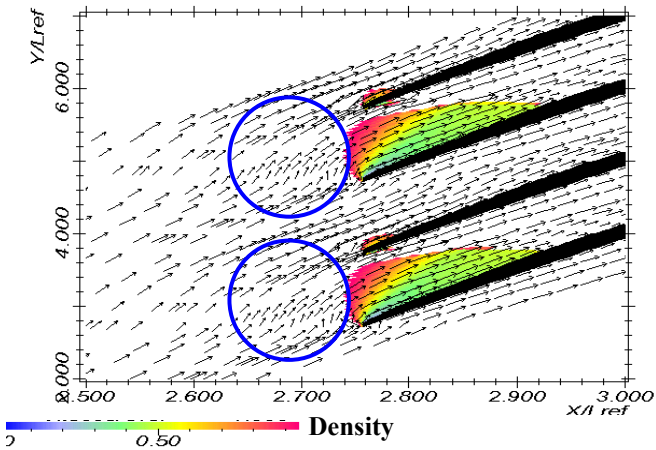

**Fig. 10.** Deviation of velocities due to high void ratio ($T=110T_{ref}$)

A spectral analysis of the mass flow rate evolution downstream from the blades is proposed in Fig. 11a. The frequencies are indicated in the rotating frame. A characteristic frequency $f_c = 0.16\ f_0$ ($f_0$ is the rotating frequency of the runner) is obtained, which is consistent with the periodical mass flow rate evolutions reported previously in Fig. 8. $f_c$ thus corresponds to the evolution of the sheets of cavitation on the blades. This phenomenon also clearly appears on the spectral analysis in the absolute frame of the static pressure far upstream from the blades, drawn in Figure 11b: the frequency $f_{ca} = 0.32\ f_0$ is predominant. $f_{ca} = 2 \times f_c$, the factor 2 being due to the alternate cavitation pattern on the blades. We will name this configuration sub-synchronous rotating cavitation, since $f_{ca}$ is lower than $f_0$.

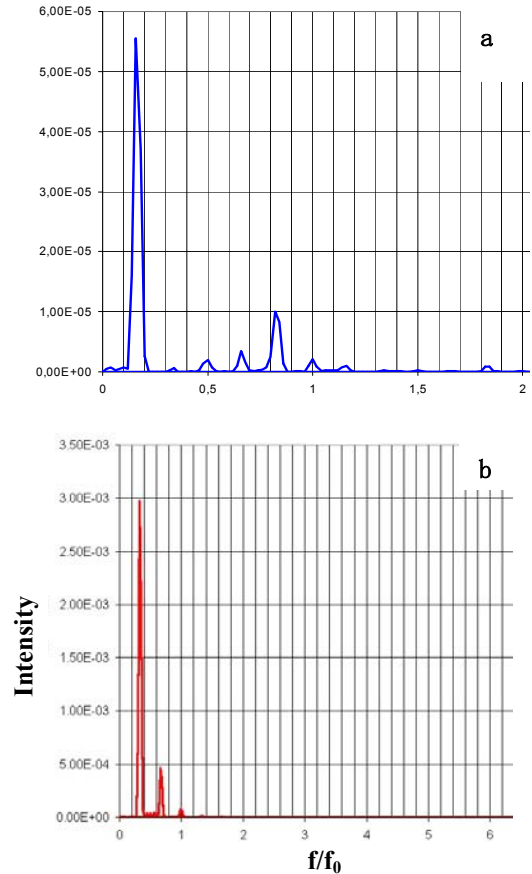

**Fig. 11.** Spectral analysis ($\sigma = 0.7$, $Q = Q_n$)
a. Mass flow rate downstream from the blades, in the rotating frame
b. Static pressure far upstream from the blades, in the absolute frame

*3.3.2 Second configuration for σ = 0.625 (nominal flow rate)*

If the cavitation number is decreased down to a value between 0.6 and 0.65, another type of unstable behaviour is obtained. It is characterised by four different sizes of the sheet of cavitation in the four channels. No correlation between channels is detectable at any time, as can be seen in Fig. 12, in which the evolution during 155 $T/T_{ref}$ of the four mass flow rates in the four channels is reported. As previously, the instability is periodical after the initial transient period, with mass flow rate fluctuations of constant magnitude in all channels. Nevertheless, the mechanisms that govern the cavity evolution in each channel are much more complicated than in the first unstable configuration, since all cavities interact.

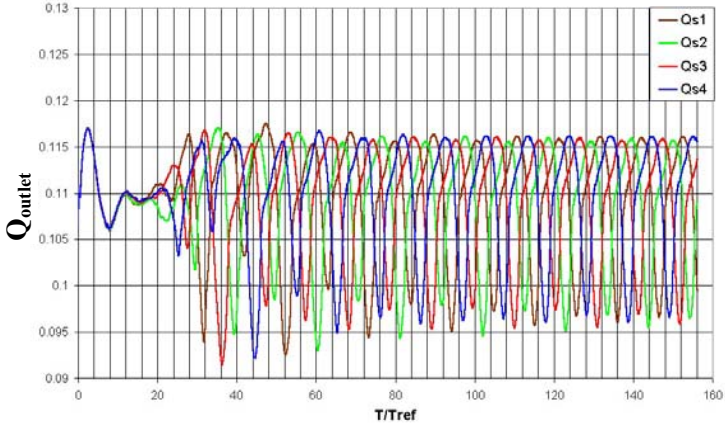

**Fig. 12.** Time evolution of the mass flow rates in the four channels (σ = 0.625, Q = $Q_n$)

The spectral analysis of the mass flow rate downstream from the blade is proposed in Fig. 13a. A characteristic frequency $f_c = 0.38\ f_0$ clearly appears in the rotating frame, with two harmonic frequencies. It corresponds to the evolution of the mass flow rates in the four channels, and it indicates that this phenomenon in the absolute frame has a higher frequency than the runner rotation ($f_c = 1.38$ in the absolute frame). This is the reason why such unstable configuration is called super-synchronous rotating cavitation. It means that the configuration in the runner at a given time (the four sizes of the cavities) rotates faster than the runner itself. The characteristic frequency in the absolute frame is obtained in Fig. 13b., in the case of the spectral analysis of the upstream static pressure evolution. The influence of the blade passage also slightly appears at $f_0$ and compounds of $f_0$.

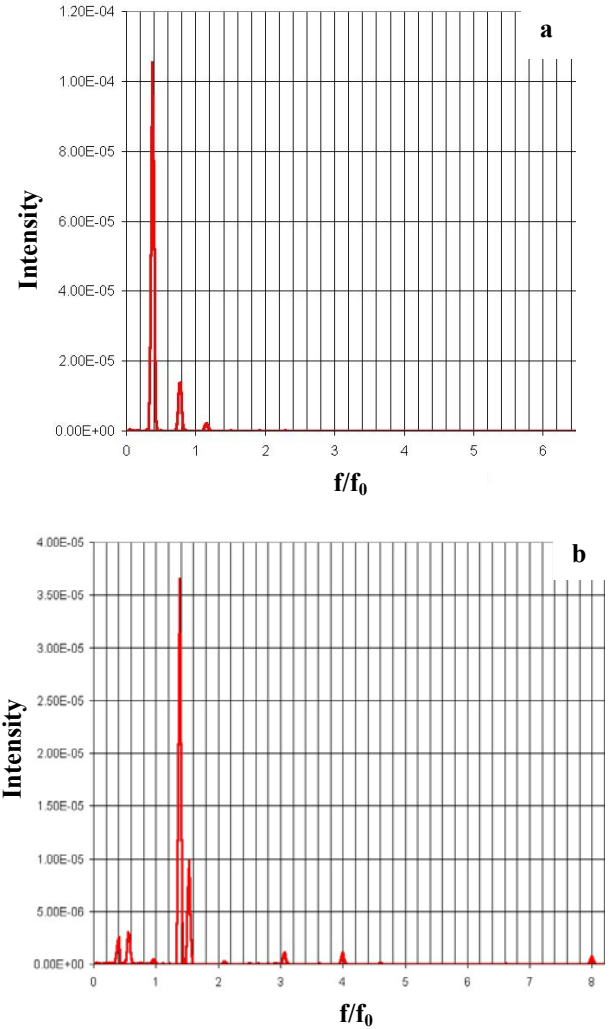

**Fig. 13.** Spectral analysis (σ = 0.625, Q = $Q_n$)

a. Mass flow rate downstream from the blades, in the rotating frame
b. Static pressure far upstream from the blades, in the absolute frame

**3.4. Efforts on the blade in unstable configurations**

*3.4.1 Sub-synchronous configuration*

To improve the understanding of the effects of rotating cavitation on the blade and on the entire runner, the fluctuations of the forces acting on the blades have been calculated and drawn in Fig. 14. The objective is to estimate the magnitude of the stress fluctuations. The stresses are calculated through the momentum conservation in a control volume surrounding one blade. Projecting this relation respectively on the normal to the inlet boundary and on the normal to the rotation axis gives the axial and the transverse components of the non-dimensional force, whose time evolution is then represented in. Fig. 14a. and 14b. for three different blades. Stress fluctuations in these figures are generated by rotating cavitation.

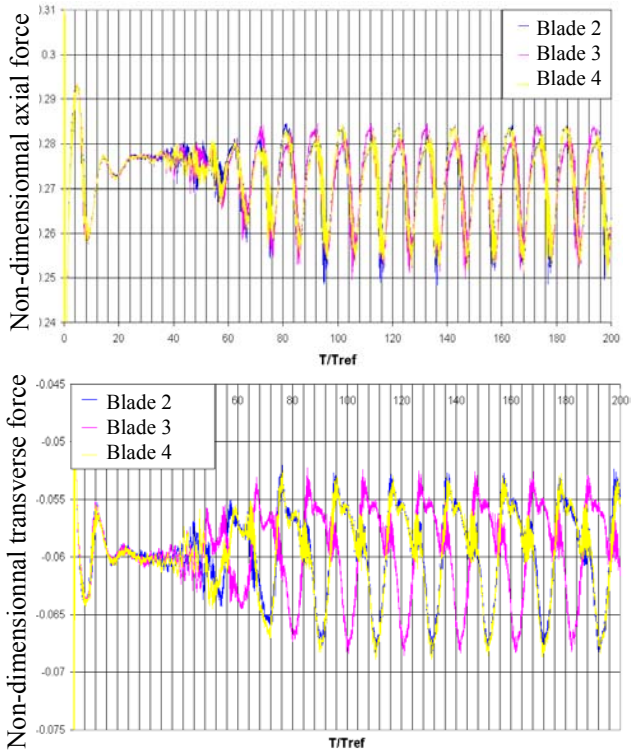

**Fig. 14.** Time distribution of the axial and transverse forces on the blades ($\sigma = 0.7$ and $Q = Qn$)

The mean non-dimensional axial force $Fx/(\rho_{ref}V_{ref}^2 L_{ref}^2)$ can be estimated to 0.27 whatever the blade we consider, and the order of magnitude of the fluctuations is about 0.016. Thus, rotating cavitation leads in the present case to fluctuations of the axial force equal to 6% of the mean force. In the case of the transverse force, the fluctuations reach 13% of the mean force.

*3.4.2 Super-synchronous configuration*

The corresponding force evolutions are represented in Fig 15a and 15b. In the case of the second blade for example, the mean axial force still equals 0.27, while the fluctuations magnitude is about 0.03: rotating cavitation is responsible in this case for fluctuations whose magnitude can be estimated to 10% of the mean effort. In the case of the transverse force, the fluctuations reach almost 20% of the mean force.

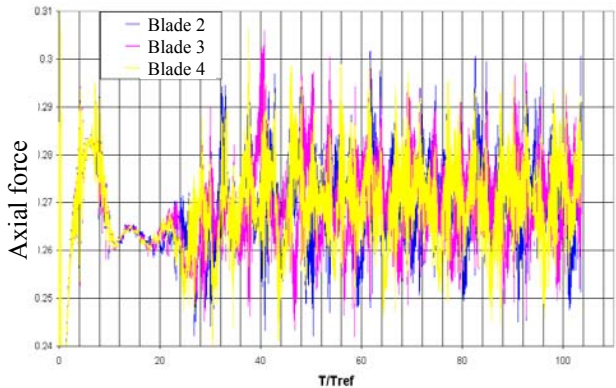

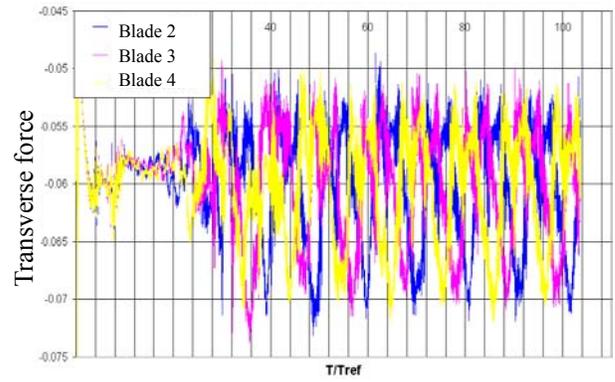

**Fig. 15.** Axial and transverse forces on the blades ($\sigma = 0.625$ and $Q = Qn$)

According to these first calculations, the simulated sub-synchronous rotating cavitation seems to be less critical for the runner than with the calculated super-synchronous rotating cavitation. In this case, axial forces on the blades substantially increase, and the periodical forces acting on the blades could be responsible for some damage to the inducer.

## 4. CONCLUSION

Computations were performed at different operating conditions by varying the mass flow rate and the cavitation number. The resulting performance charts of the blade cascade at the calculated mass flow rates are reported in Fig. 16. Stable and unstable cavitation configurations are also indicated, so that the limit of the rotating cavitation behaviour can be identified. Results for mass flow rates lower than 0.9 Qn are not reported because of the numerical difficulties that we encountered at these flow conditions. Indeed, at partial flow rate, incidence of the flow at the blade leading edge and cavity width both increase. In case of rotating behaviour, it can result in the complete obstruction of a channel, which induces serious numerical problems.

It can be seen that the cavitation parameter range of rotating cavitation increases when the flow rate coefficient decreases. That result agrees with experimental observations [Pagnier et al. 1995, Yokata et al., 1999]. At reference flow rate Qn the experimental instability range is about $\Delta\sigma_{exp}=$ 0.15, which is consistent with the results obtained previously by [Joussellin et al., 2001] in the case of a similar four-blade inducer geometry. The mass flow rate limit for appearance of unstable configuration (1.2 Qn) is also identical to the one obtained in this previous study. Further work is now needed to assess the prediction capability of the model. Therefore, improvements of the analysis of 3D inlet flows and study of the influence of the cascade design by numerical simulations are in progress, in parallel to the development of a full 3D model [Coutier-Delgosha et al., 2002].

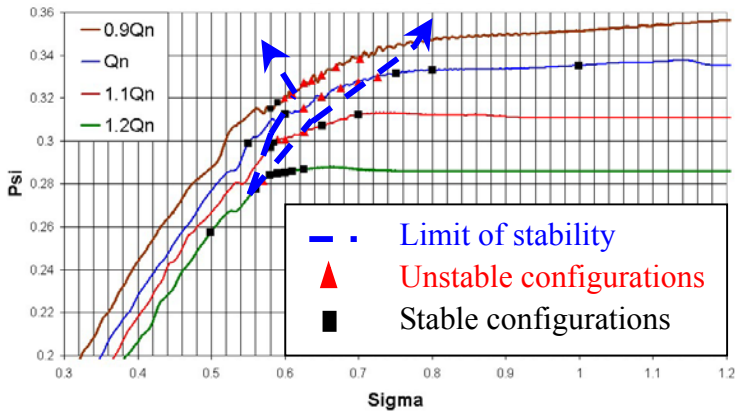

**Fig. 16.** Performance charts at several mass flow rates

## ACKNOWLEDGMENTS

The authors wish to thank SNECMA Moteurs (Rocket Engine Division) and the French space agency CNES for their support to the present work.